\documentclass[12pt]{article}
\usepackage{graphicx}

\newcommand\pubnumber{ATL-PHYS-PROC-2016-252}
\newcommand\pubdate{December 2, 2016}

\def\institute{Nagoya University, JAPAN}
\def\instituteB{University of Illinois at Urbana-Champaign, United States}

\def\Title#1{\begin{center} {\Large #1 } \end{center}}
\def\Author#1{\begin{center}{ \sc #1} \end{center}}
\def\Address#1{\begin{center}{ \it #1} \end{center}}

\newcommand\pubblock{\rightline{\begin{tabular}{l} \pubnumber\\
         \pubdate  \end{tabular}}}
\newenvironment{Abstract}{\begin{quotation}  }{\end{quotation}}
\newenvironment{Presented}{\begin{quotation} \begin{center} 
             PRESENTED AT\end{center}\bigskip 
      \begin{center}\begin{large}}{\end{large}\end{center} \end{quotation}}

\def\beq{\begin{equation}}
\def\eeq#1{\label{#1}\end{equation}}
\def\eeqn{\end{equation}}

\def\beqa{\begin{eqnarray}}
\def\eeqa#1{\label{#1}\end{eqnarray}}
\def\eeqan{\end{eqnarray}}

\let\bar=\overbar

\def\Dslash{\not{\hbox{\kern-4pt $D$}}}
\def\dslash{\not{\hbox{\kern-2pt $\del$}}}

\def\msb{{\bar{\ssstyle M \kern -1pt S}}}

\begin{document}
\begin{titlepage}
\pubblock

\vfill
\Title{Measurement of top quark pair differential cross-sections in the dilepton channel in $pp$ collisions at $\sqrt{s} =$ 7 and 8\,TeV with ATLAS}
\vfill
\Author{Kentaro Kawade}
\Address{\institute}
\Author{Ki Lie}
\Address{\instituteB}
\Author{on behalf of the ATLAS Collaboration}
\vfill
\begin{Abstract}
Measurements of normalized differential cross-sections of top quark pair ($t\bar t$) production are presented as a function of the mass,
the transverse momentum and the rapidity of the $t\bar t$ system in proton-proton collisions at center-of-mass energies of $\sqrt{s}$ = 7 TeV and 8 TeV.
The dataset corresponds to an integrated luminosity of 4.6 fb$^{-1}$ at 7 TeV and 20.2 fb$^{-1}$ at 8 TeV, recorded with the ATLAS detector at the Large Hadron Collider. Events with top quark pair signatures are selected in the dilepton final state, requiring exactly two charged leptons and at least two jets with at least one of the jets identified as likely to contain a $b$-hadron. The measured distributions are corrected for detector effects and selection efficiency to cross-sections at the parton level. The differential cross-sections are compared with different Monte Carlo generators and theoretical calculations of $t\bar t$ production. The results are consistent with the majority of predictions in a wide kinematic range.
\end{Abstract}
\vfill
\begin{Presented}
$9^{th}$ International Workshop on Top Quark Physics Olomouc, Czech Republic,  September 19--23, 2016
\end{Presented}
\vfill
\end{titlepage}
\def\thefootnote{\fnsymbol{footnote}}
\setcounter{footnote}{0}

\section{Introduction}

The top quark is the most massive elementary particle in the Standard Model (SM).
The production of top quarks at the Large Hadron Collider (LHC) is dominated
by pair production of top and antitop quarks ($t\bar{t}$) via the strong interaction,
and thus measurements of these distributions provide a means of testing the SM prediction at the TeV scale.
Since new phenomena beyond the SM can modify the kinematic properties of the $t\bar{t}$ system,
and $t\bar{t}$ events are the dominant background to many searches for new physics,
more accurate and detailed knowledge of top quark pair production is an essential component of the wide-ranging LHC physics program.

In this paper, the measurements of normalized differential top quark pair production cross-sections
are presented as a function of the mass ($m_{t\bar{t}}$), the transverse momentum ($p_{T,t\bar{t}}$) and the rapidity
($y_{t\bar{t}}$) of the $t\bar{t}$ system in 
proton-proton collisions at center-of-mass energies of $\sqrt{s} =$ 7\,TeV and 8\,TeV.
The data-set corresponds to an integrated luminosity of 4.6\,${\rm fb^{-1}}$ at 7\,TeV and 20.2\,${\rm fb^{-1}}$ at 8\,TeV,
recorded with the ATLAS detector \cite{PERF-2007-01} at the LHC.

\section{Analysis}
Events with top quark pair signatures are selected in the dilepton final state, in which both $W$ bosons from top quarks decay leptonically,
requiring exactly two charged leptons and at least two jets where at least one of them identified as likely to contain a $b$-hadron.
In the 8\,TeV measurement, only the $e\mu$ channel is considered, 
while in the 7\,TeV analysis, where the integrated luminosity is smaller,
events containing same-flavor electron or muon pairs (the $ee$ and $\mu\mu$ channels) are also selected
to maximize the size of the available data-set.

\begin{figure}[htb]
\centering
\includegraphics[height=1.85in]{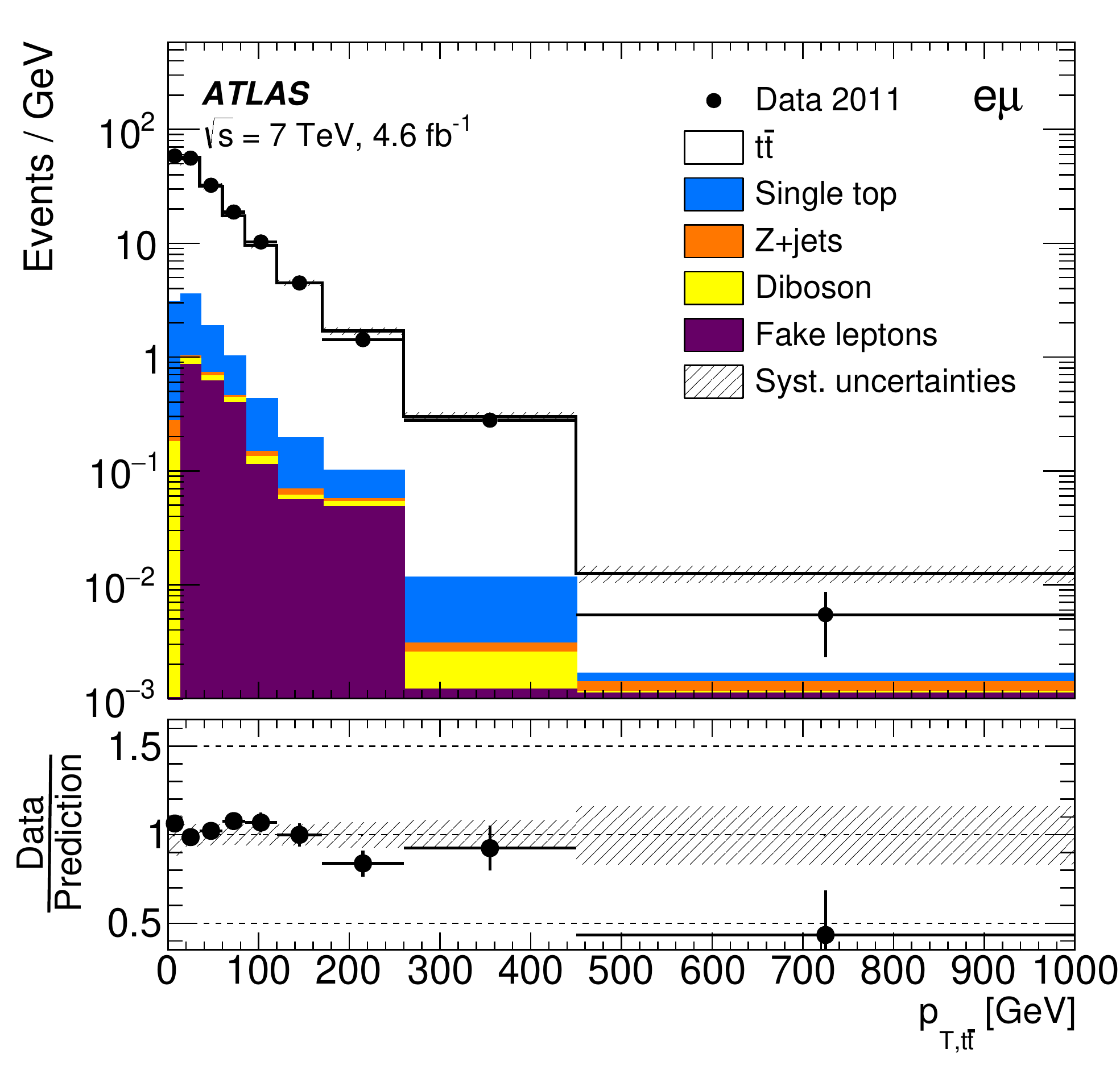}
\includegraphics[height=1.85in]{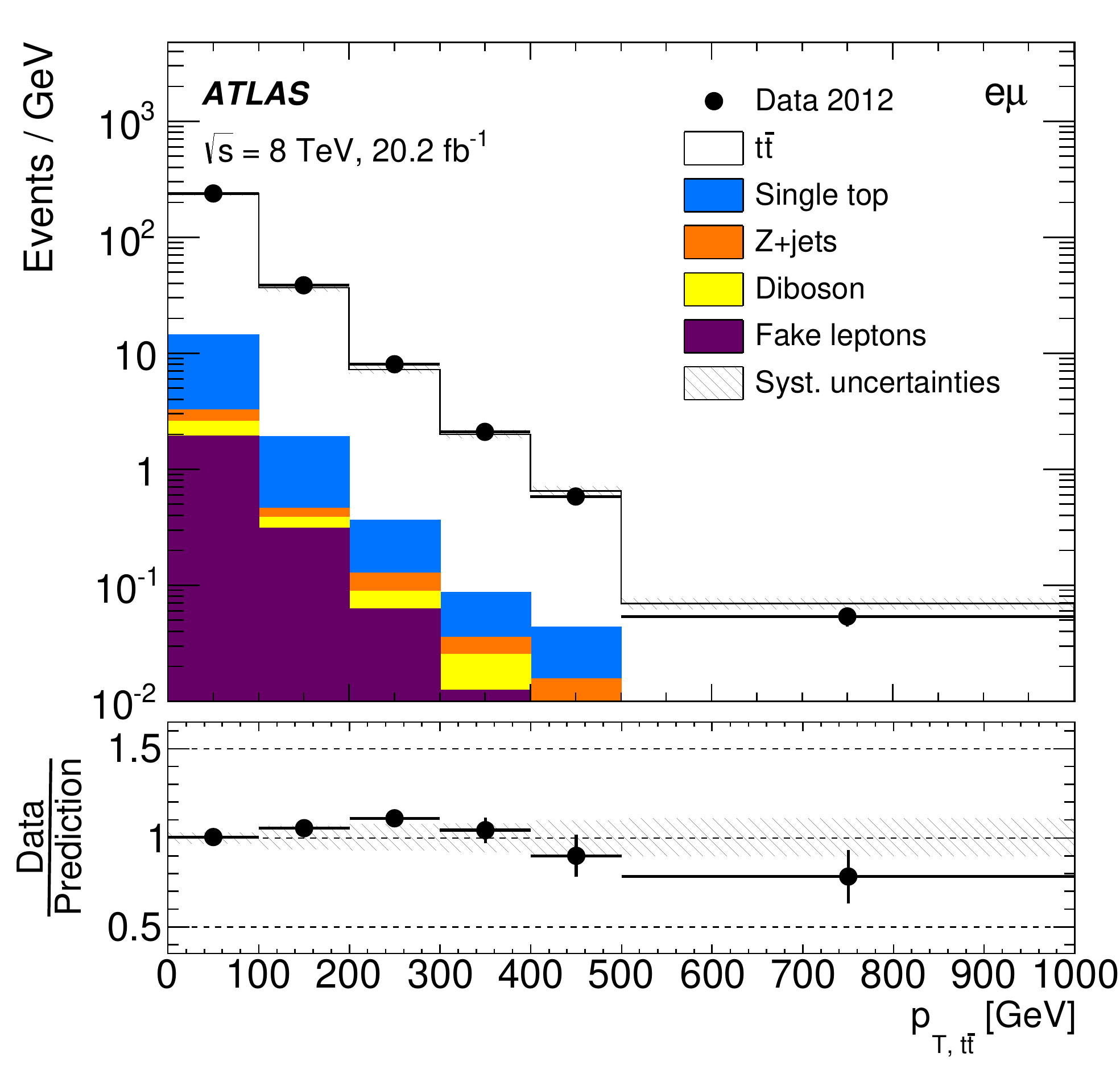}
\caption{The reconstructed distributions of $p_{T,t\bar{t}}$ of the $t\bar{t}$ system obtained from the $\sqrt{s}$ = 7 and 8\,TeV
  data compared with the total signal and background predictions, in the $e\mu$ channels \cite{Journal}.
  The bottom panel shows the ratio of data to prediction. 
  The error band includes all systematic uncertainties except $t\bar{t}$ modeling uncertainties. 
}
\label{fig:recoplot}
\end{figure}

An approximate four-momentum of the $t\bar{t}$ system is reconstructed from two leptons, two jets, and missing transverse momentum.
Only leptons passing good quality criteria and the two jets with highest probability to contain a $b$-hadron are considered.
Figure \ref{fig:recoplot} shows the reconstructed $p_{T,t\bar{t}}$ distributions together with the MC predictions at 7\,TeV and 8\,TeV.
Overall there is good agreement between data and prediction.

\begin{figure}[h]
\centering
\includegraphics[height=1.85in]{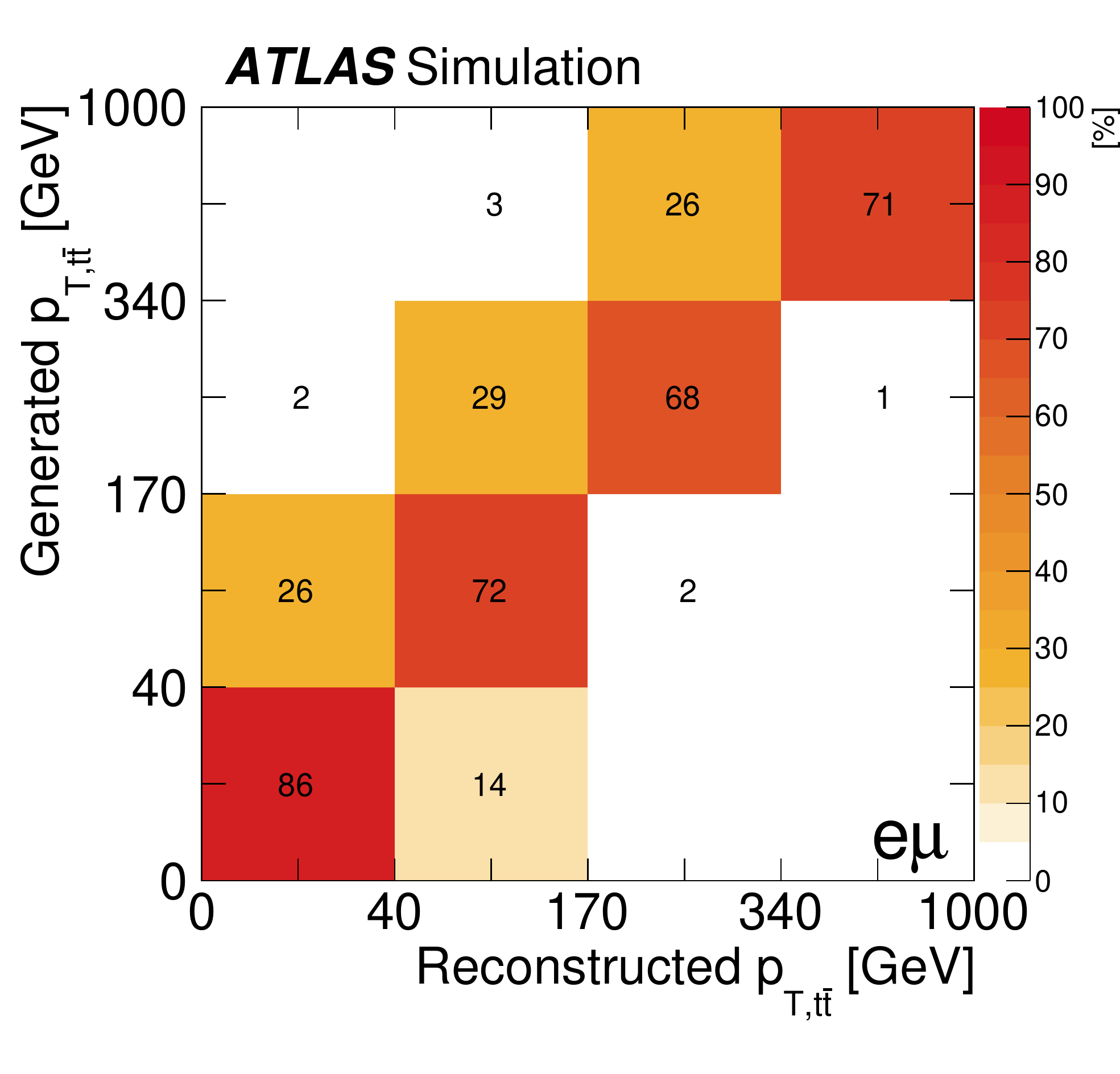}
\includegraphics[height=1.85in]{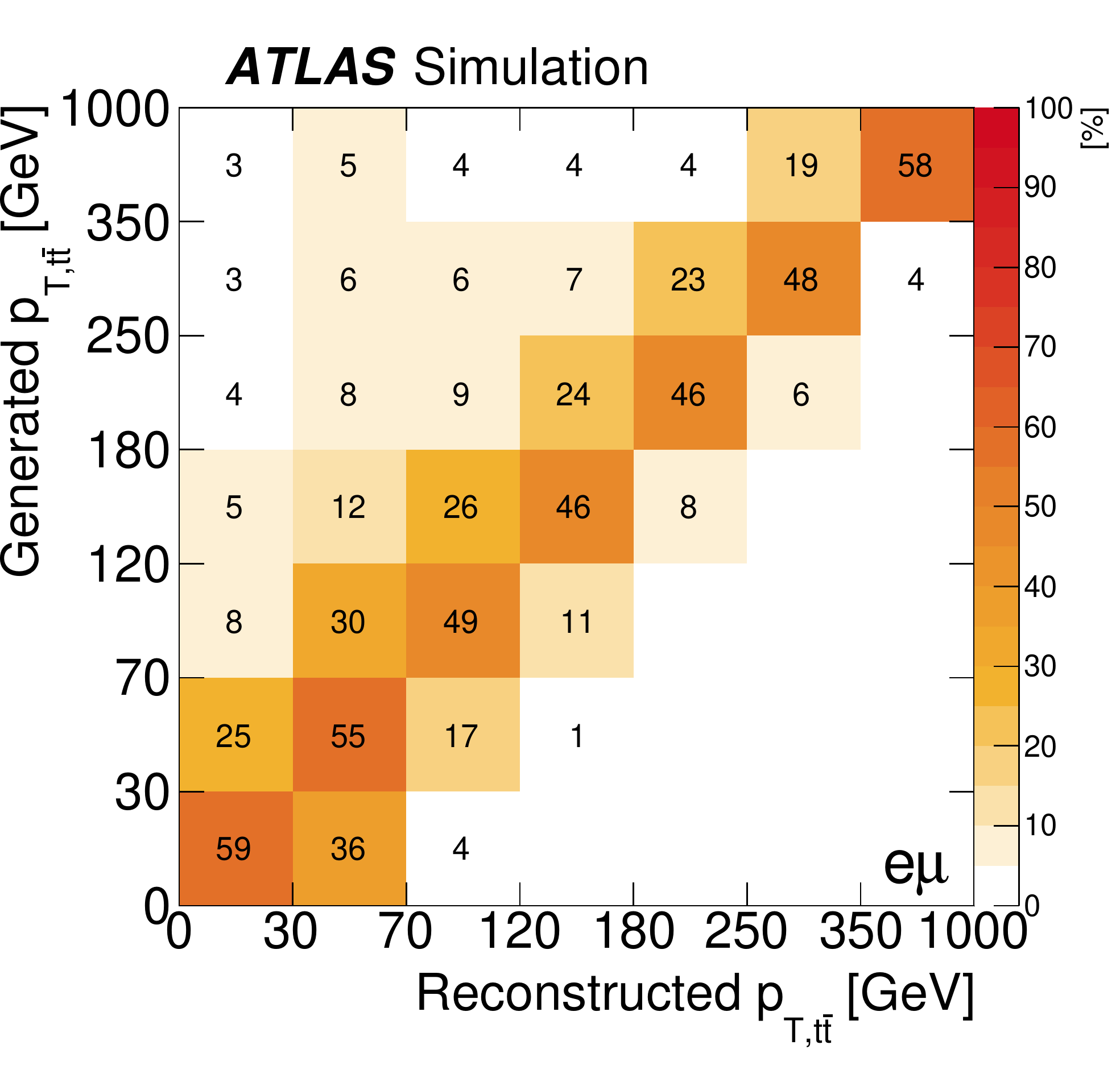}
\caption{The migration matrix of $p_{T,t\bar{t}}$ represented in probability for 7\,TeV (left) and 8\,TeV (right)
  in the $e\mu$ channel, obtained from $t\bar{t}$ simulation with the {\sc Powheg+Pythia} generator \cite{Journal}.
  Elements in each row add up to unity.}
\label{fig:response}
\end{figure}

After subtraction of the estimated backgrounds, the measured distributions are corrected for detector effects
and selection efficiency to cross-sections at the parton level by employing iterative Bayesian technique,
with the migration matrix determined using the $t\bar{t}$ MC simulations with {\sc Powheg+Pythia}.
Figure \ref{fig:response} presents the migration matrices of $p_{T,t\bar{t}}$ for both 7\,TeV and 8\,TeV in the $e\mu$ channel.
The matrix element $M_{ij}$ represents the probability for an event generated at parton level in bin $i$ to be
 reconstructed in bin $j$, so the elements of each row add up to unity.

The uncertainties from the signal modeling and detector modeling, which affect the estimation of the detector response
and the signal reconstruction efficiency, are estimated by repeating the full analysis procedure with each systematic sample.
The uncertainties from the signal modeling and the jet energy scale calibration are dominant for 
$m_{t\bar{t}}$ and $p_{T,t\bar{t}}$ measurements,
while the uncertainty due to PDF is dominant for $y_{t\bar{t}}$ measurement.
The uncertainty due to the background estimation is based on both data and MC simulation and 
mainly visible in higher $m_{t\bar{t}}$ and $p_{T,t\bar{t}}$ regions.

\section{Results and discussion}
\begin{figure}[h]
\centering
\includegraphics[height=1.85in]{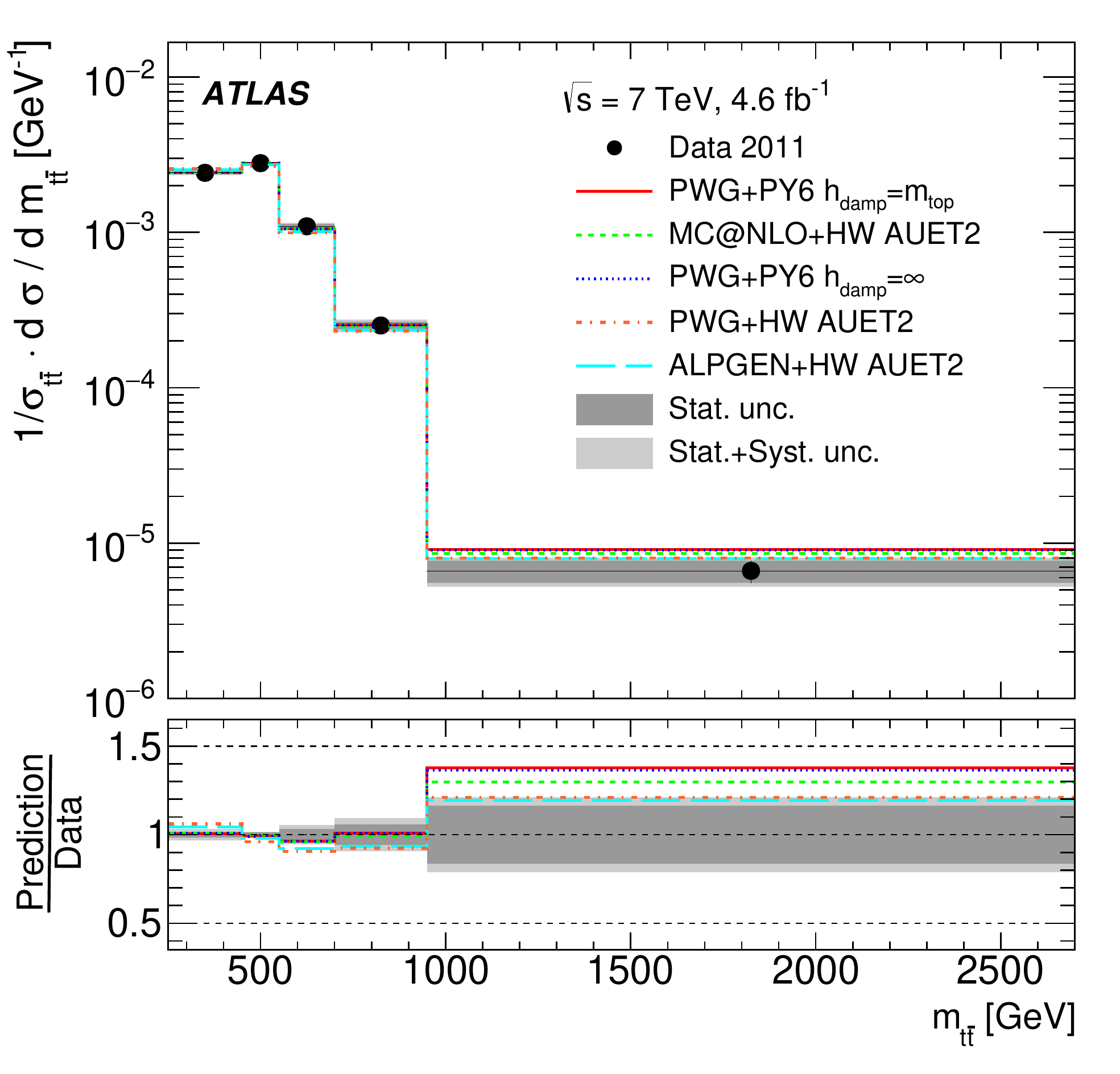}
\includegraphics[height=1.85in]{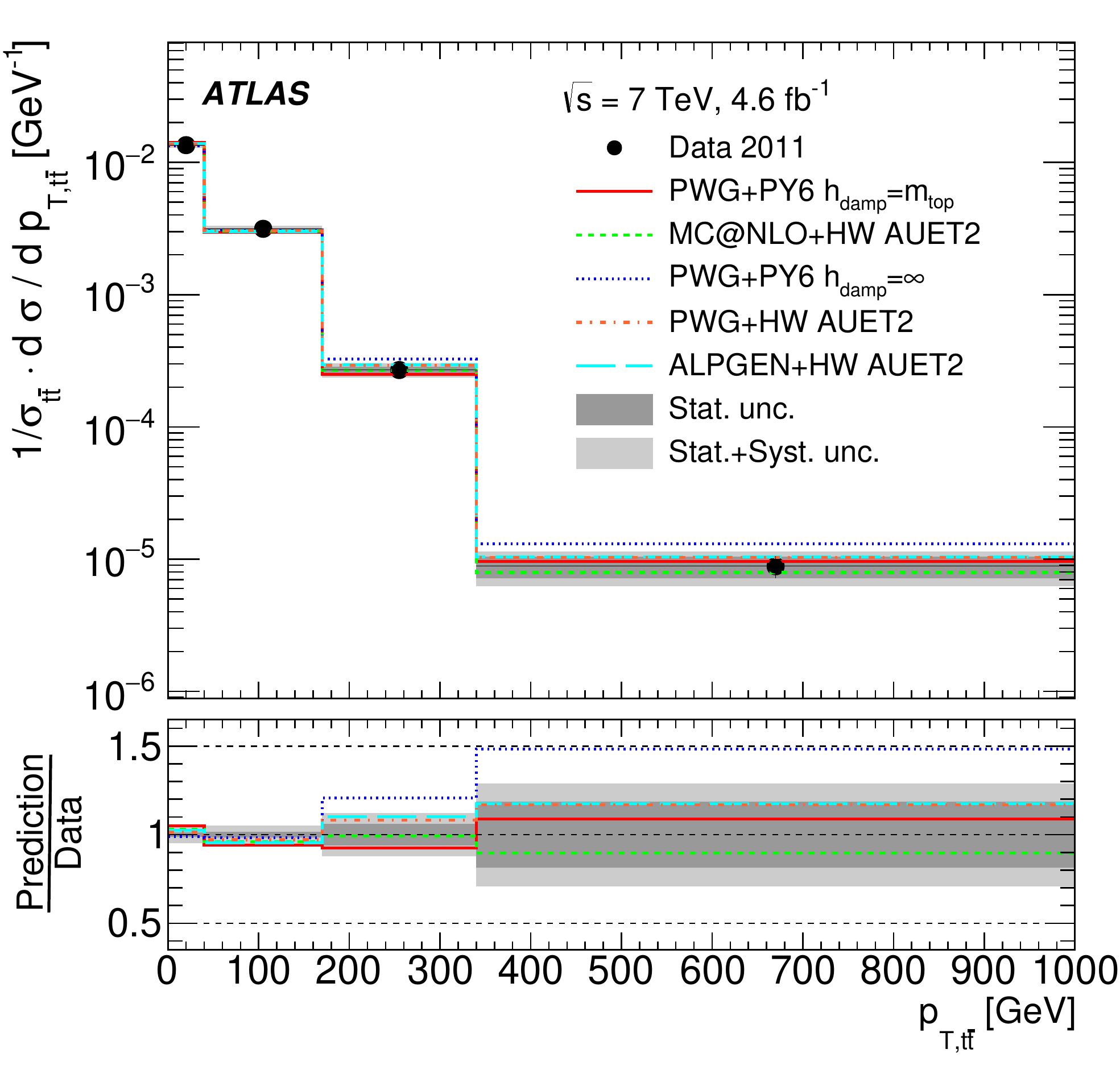}
\includegraphics[height=1.85in]{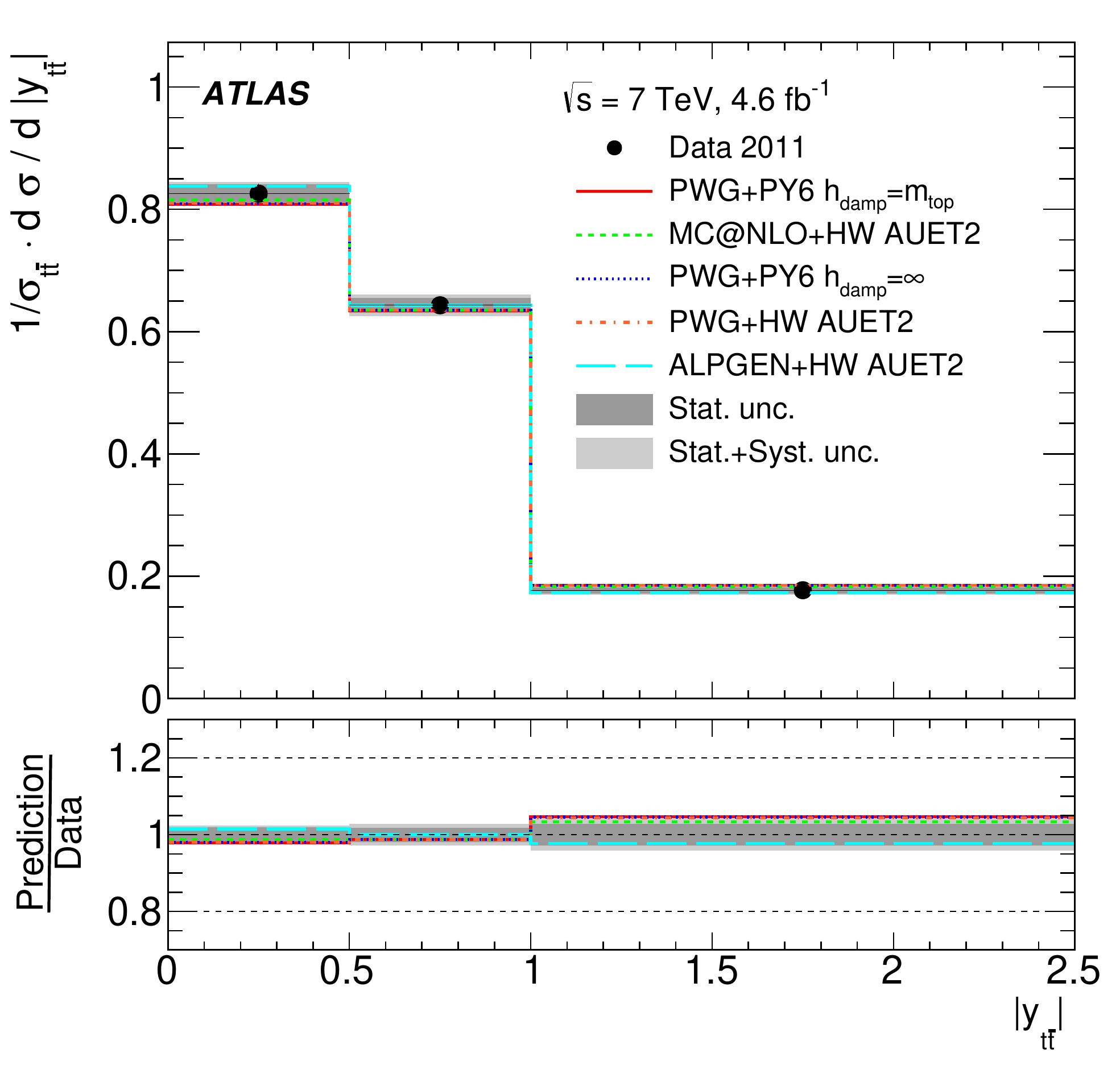}
\caption{Normalized $t\bar{t}$ differential cross-sections as a function of $m_{t\bar{t}}$, $p_{T,t\bar{t}}$
  and $|y_{t\bar{t}}|$ at $\sqrt{s} =$ 7\,TeV together with the MC predictions \cite{Journal}.
  The bottom panel shows the ratio of prediction to data.}
\label{fig:7tev}
\end{figure}

\begin{figure}[h]
\centering
\includegraphics[height=1.85in]{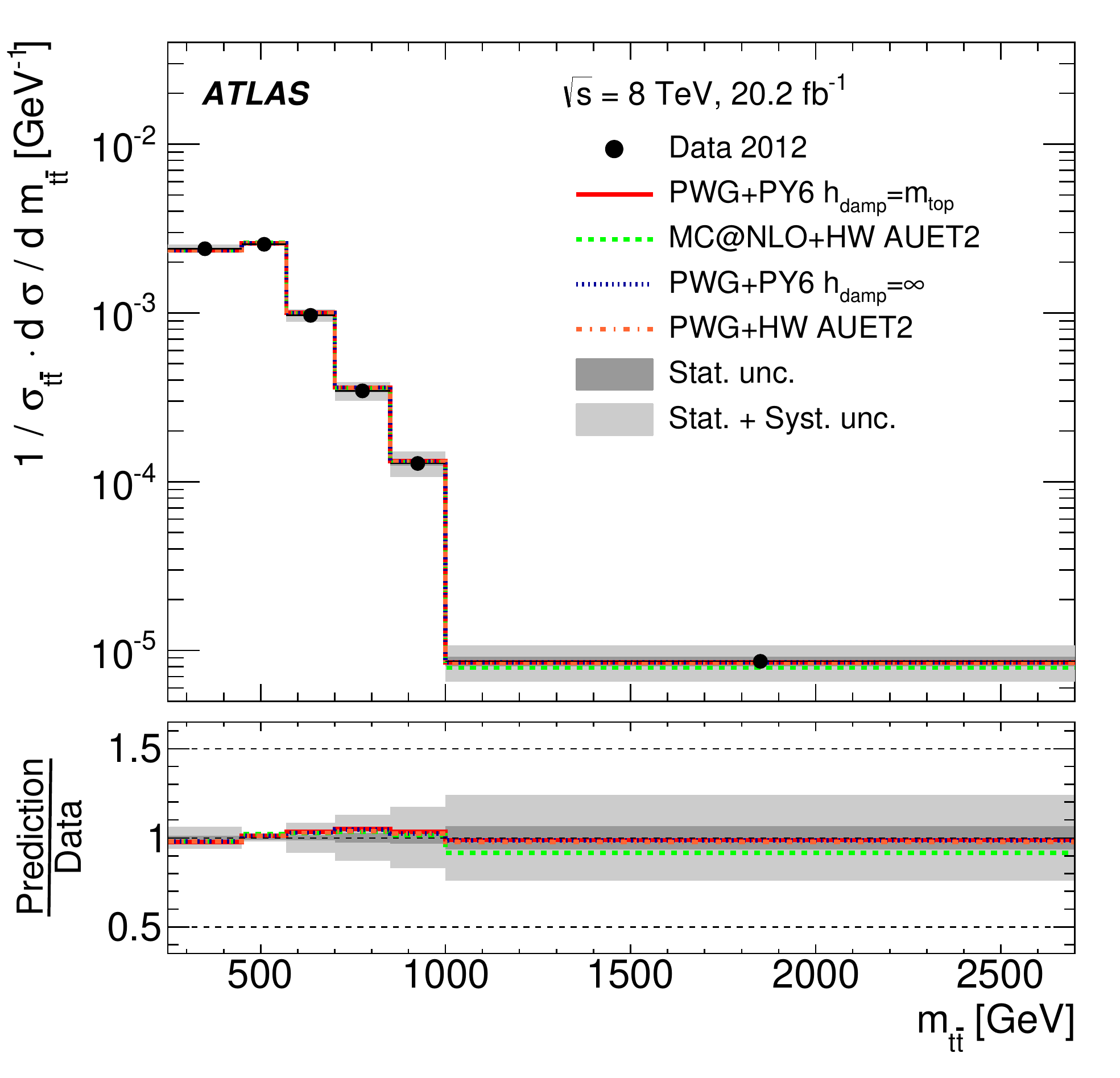}
\includegraphics[height=1.85in]{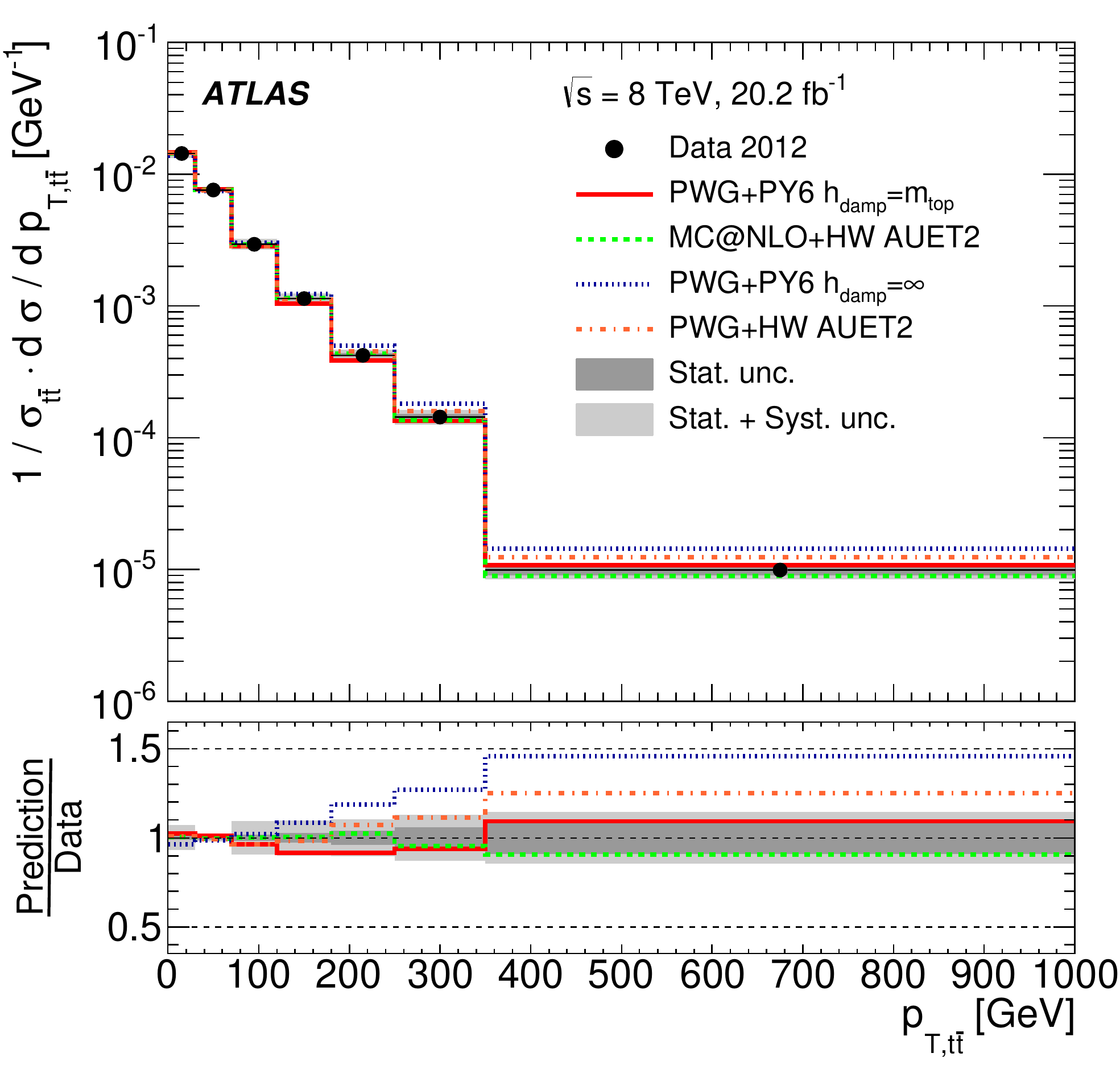}
\includegraphics[height=1.85in]{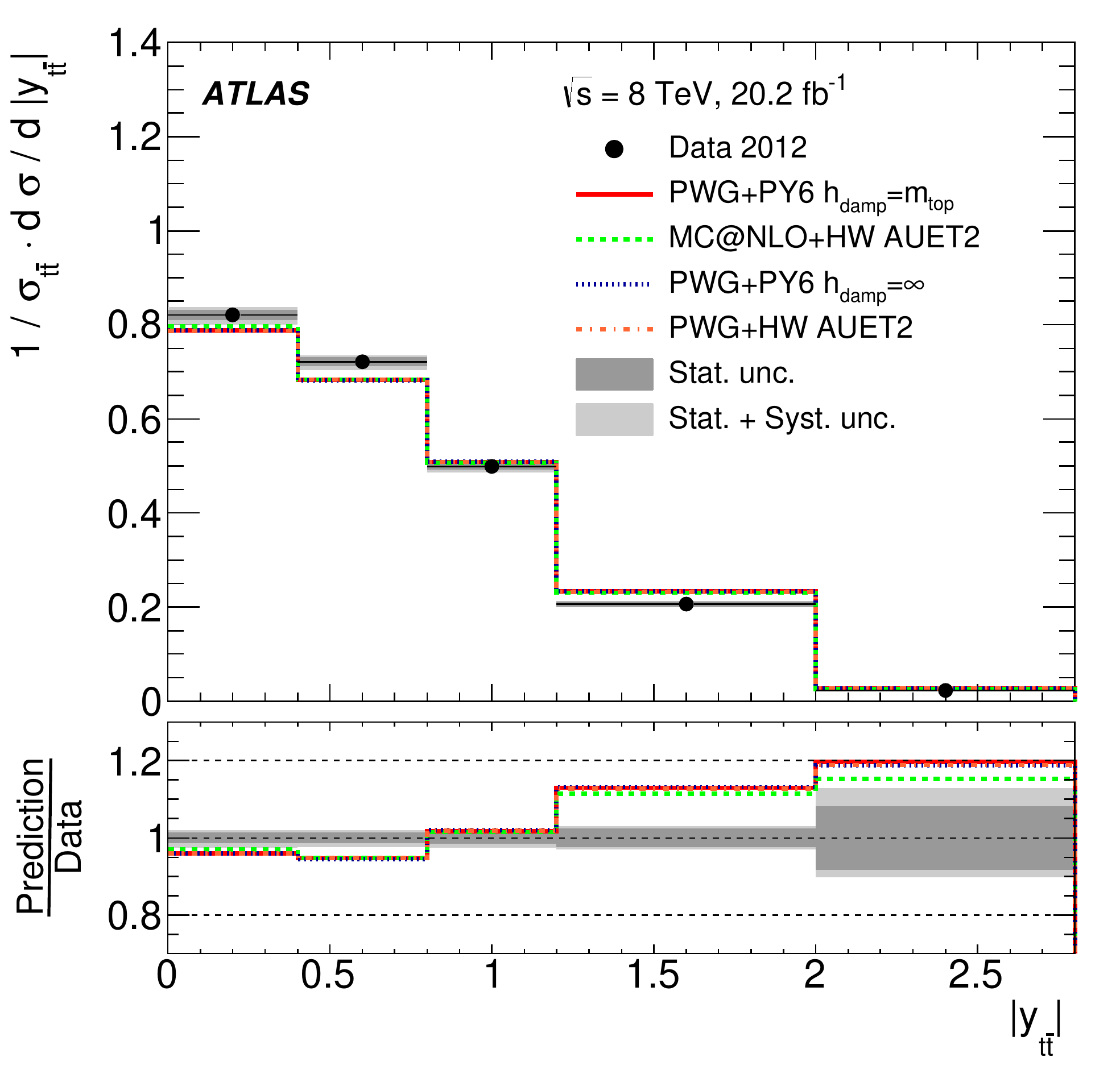}
\caption{Normalized $t\bar{t}$ differential cross-sections as a function of $m_{t\bar{t}}$, $p_{T,t\bar{t}}$
  and $|y_{t\bar{t}}|$ at $\sqrt{s} =$ 8\,TeV together with the MC predictions \cite{Journal}.
  The bottom panel shows the ratio of prediction to data.}
\label{fig:8tev}
\end{figure}

\begin{figure}[h]
\centering
\includegraphics[height=1.85in]{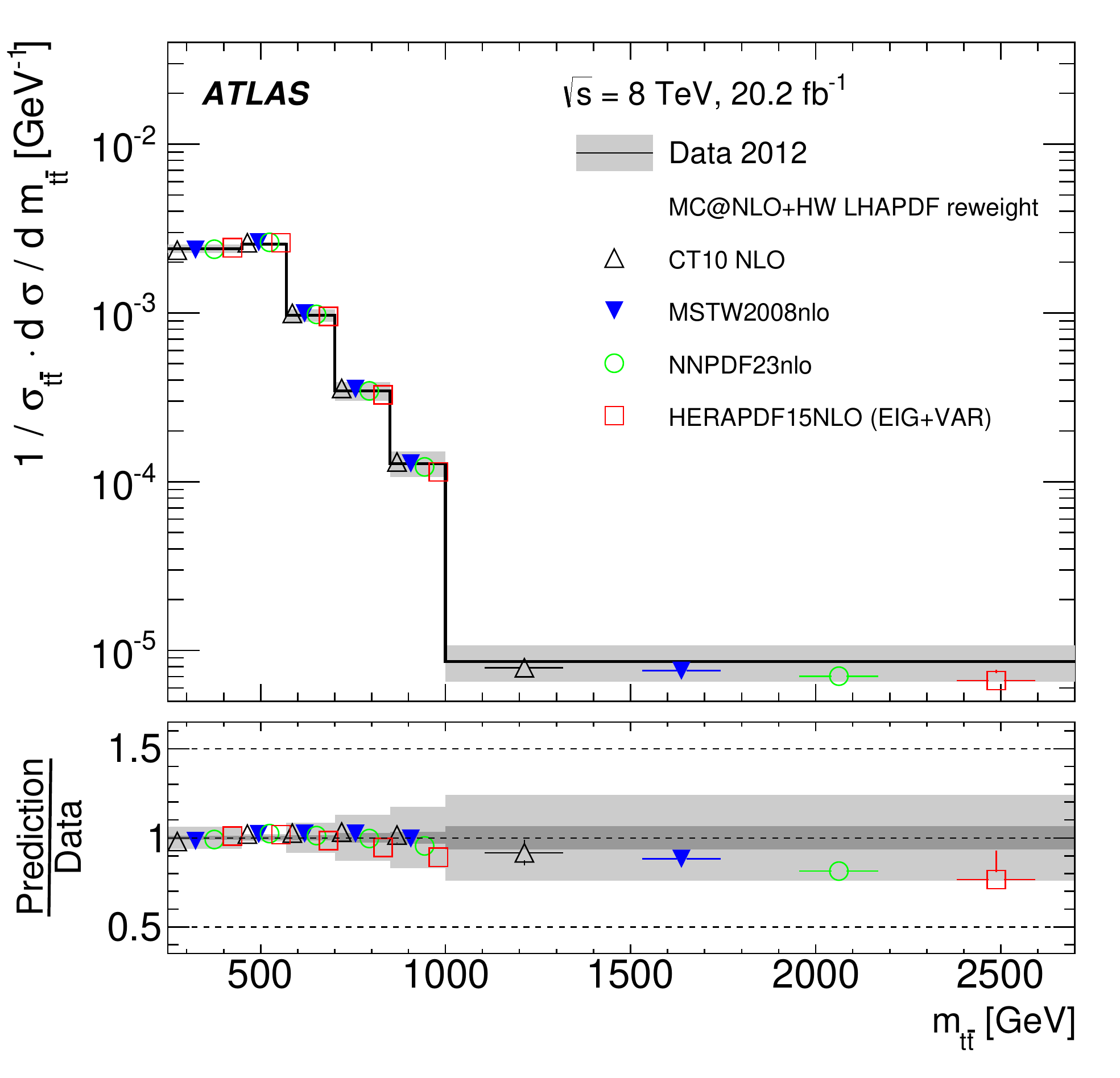}
\includegraphics[height=1.85in]{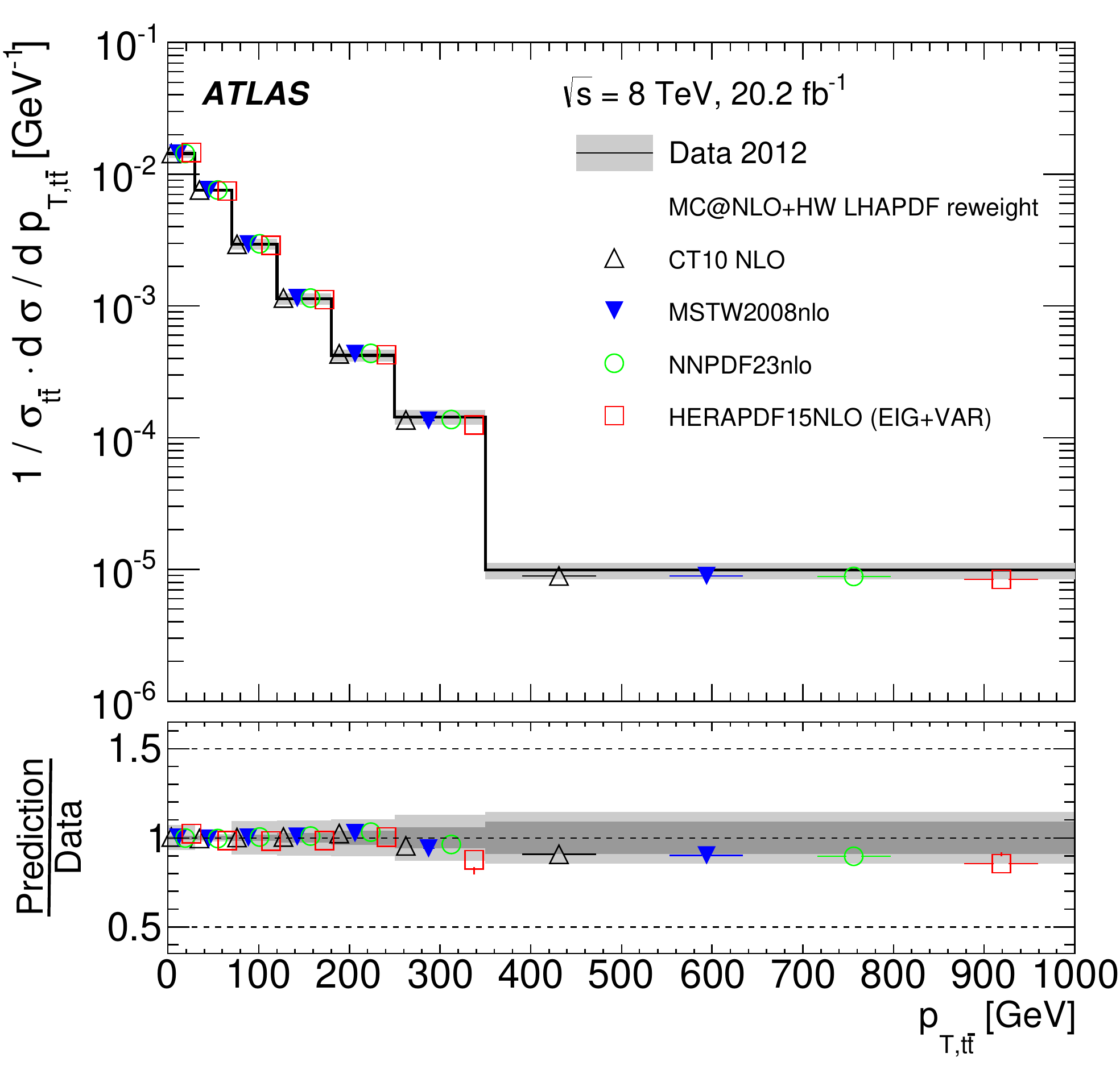}
\includegraphics[height=1.85in]{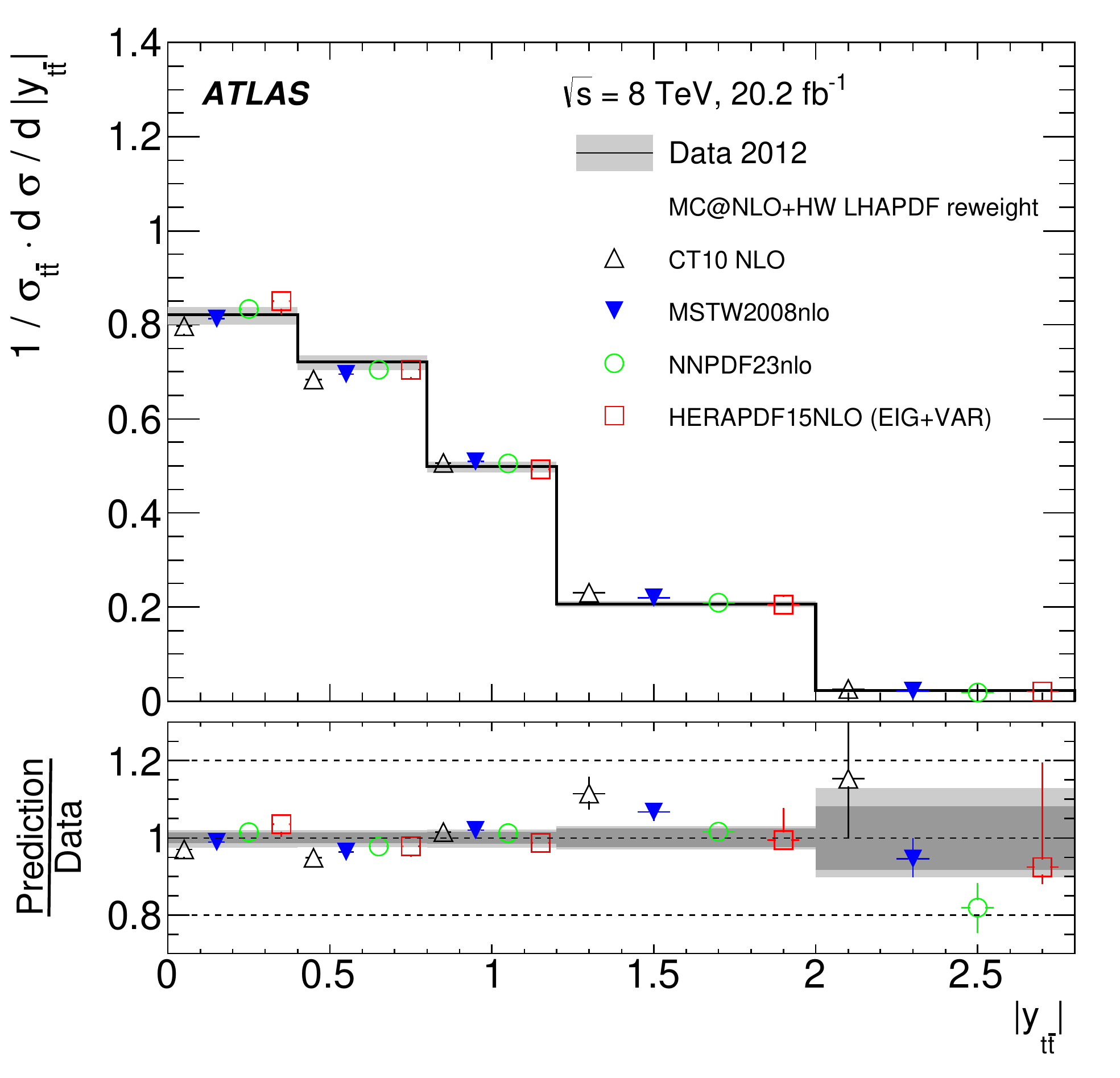}
\caption{Normalized $t\bar{t}$ differential cross-sections compared with the prediction from {\sc MC@NLO} with different PDF sets \cite{Journal}.}
\label{fig:PDF}
\end{figure}

\begin{figure}[h]
\centering
\includegraphics[height=1.85in]{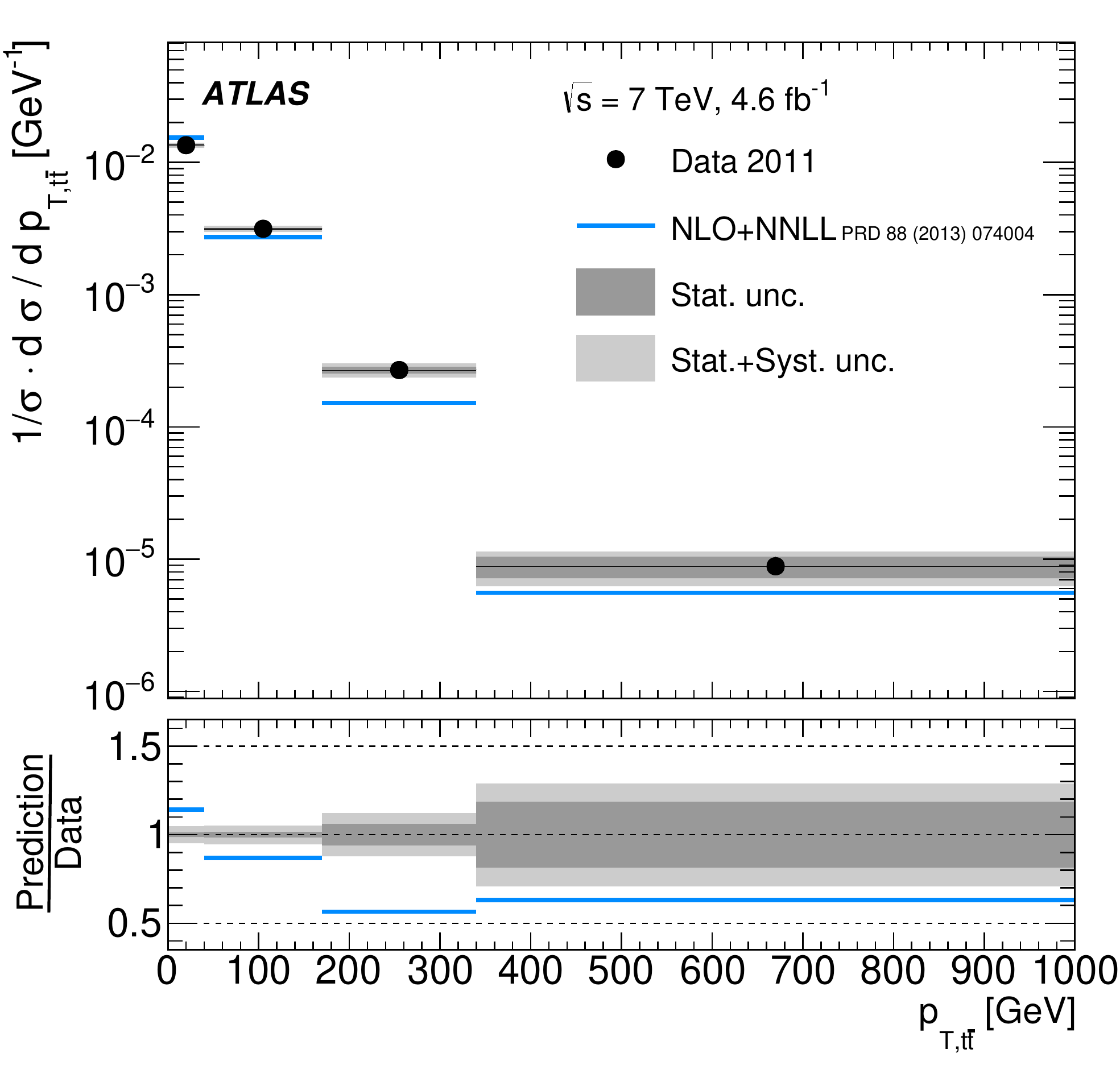}
\includegraphics[height=1.85in]{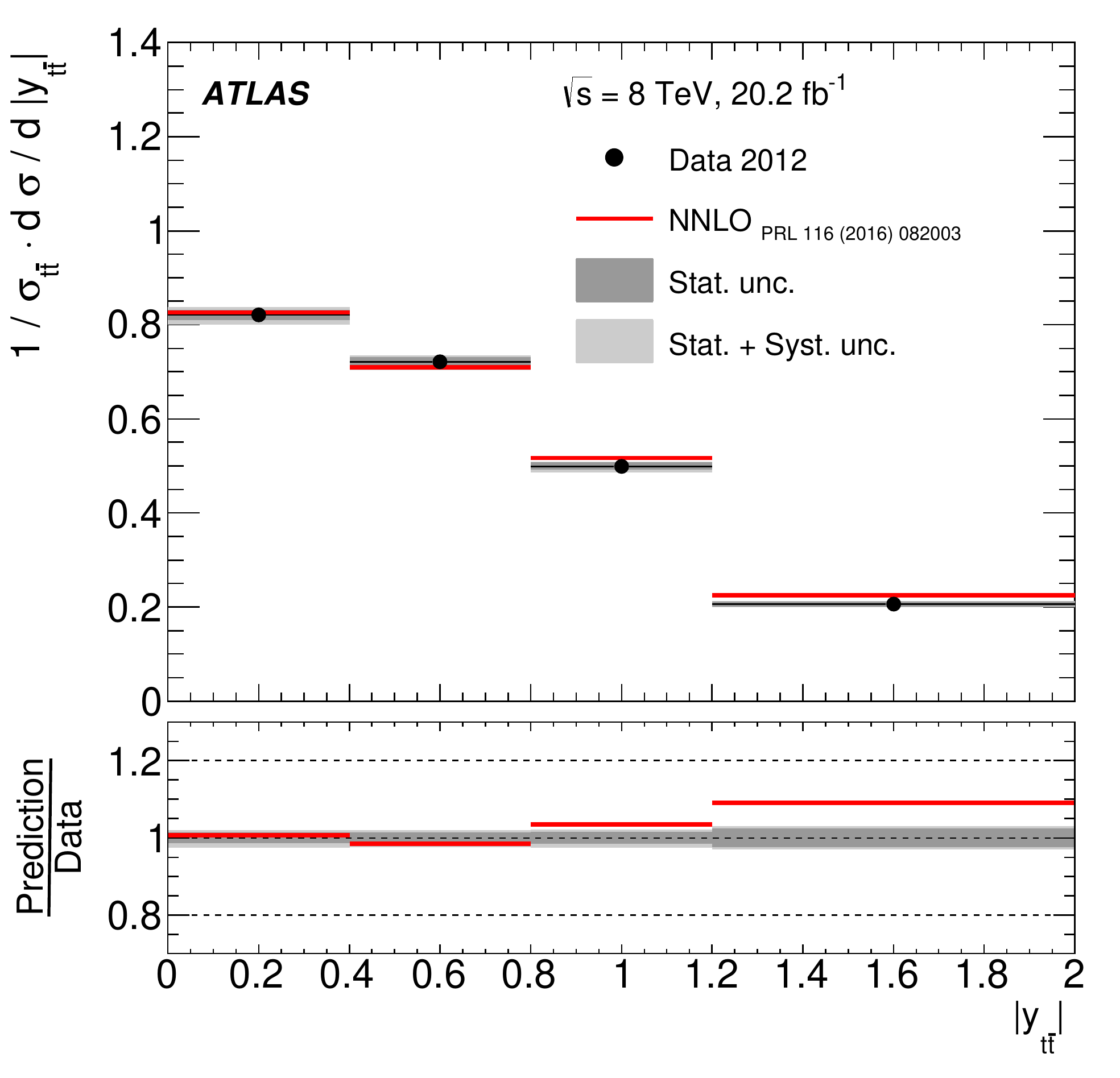}
\caption{Comparisons with NLO+NNLL prediction at 7\,TeV (left) and NNLO+NNLL prediction at 8\,TeV (right) \cite{Journal}.}
\label{fig:NNLO}
\end{figure}

Figures \ref{fig:7tev} and \ref{fig:8tev} show the comparison of the measured normalized distributions of
$m_{t\bar{t}}$, $p_{T,t\bar{t}}$, and $|y_{t\bar{t}}|$
to the predictions from different MC generators for $\sqrt{s} =$ = 7 and 8\,TeV, respectively.
The data are also compared to the MC generator with different PDF sets at as shown in figure \ref{fig:PDF},
and the higher order theoretical predictions \cite{NLO,NNLO} as shown in figure \ref{fig:NNLO}.

Most of the NLO generators agree with the experimental results in wide kinematic ranges of the distributions.
The $m_{t\bar{t}}$ spectrum is well described by most of the NLO generators at both 7\,TeV and 8\,TeV,
except for {\sc Powheg+Pythia} in the highest bin in the 7\,TeV analysis.
For $p_{T,t\bar{t}}$, agreement with {\sc Powheg+Pythia} with $h_{damp} = \inf$ is
particularly bad due to a harder $p_{T,t\bar{t}}$ spectrum than data at both 7\,TeV and 8\,TeV.
Better agreement is obtained from {\sc Powheg+Pythia} with $h_{damp} = m_t$.
In both the 7\,TeV and 8\,TeV analysis, {\sc MC@NLO+Herwig} describes the $p_{T,t\bar{t}}$ spectrum well.
For $|y_{t\bar{t}}|$, all the generators show fair agreement with data in the 7\,TeV analysis,
while at 8\,TeV, none of the generators provides an adequate description of $|y_{t\bar{t}}|$.
This difference in the level of agreement is due to the improved statistical precision and finer binning in $|y_{t\bar{t}}|$ for the 8\,TeV analysis.
More details of this work are described in Ref. \cite{Journal}.

\end{document}